# Transient-evoked otoacoustic emission signals predicting outcomes of acute sensorineural hearing loss in patients with Ménière's disease

Yi-Wen Liu[1], Sheng-Lun Kao[1], Hau-Tieng Wu[2], Tzu-Chi Liu[1], Te-Yung Fang[3,4], Pa-Chun Wang[3,4]

(1) Department of Electrical Engineering, National Tsing Hua University, Hsinchu, Taiwan

(2) Department of Mathematics and Department of Statistical Science, Duke University, Durham, North Carolina, USA

(3) Department of Otolaryngology, Cathay General Hospital, Taipei, Taiwan

(4) School of Medicine, Fu Jen Catholic University, Taipei, Taiwan

Yi-Wen Liu and Sheng-Lun Kao are joint first authors, and Te-Yung Fang and Pa-Chun Wang share equal responsibilities of this article.

**Contact Information of the Corresponding Author:**
Pa-Chun Wang, MD, MSc, MBA
Department of Otolaryngology, Cathay General Hospital
280, Sec 4. Jen-Ai Rd. 106 Taipei, Taiwan
Telephone: + 8862 2708 2121 ext 3364
FAX: + 8862 6636 2836
Email: drtony@seed.net.tw

## ABSTRACT

**Background:** Fluctuating hearing loss is characteristic of Ménière's disease (MD) during acute episodes. However, no reliable audiometric hallmarks are available for counselling the hearing recovery possibility.
**Aims/Objectives:** To find parameters for predicting MD hearing outcomes.
**Material and Methods:** We applied machine learning techniques to analyse transient-evoked otoacoustic emission (TEOAE) signals recorded from patients with MD. Thirty unilateral MD patients were recruited prospectively after onset of acute cochleo-vestibular symptoms. Serial TEOAE and pure-tone audiogram (PTA) data were recorded longitudinally. Denoised TEOAE signals were projected onto the three most prominent principal directions through a linear transformation. Binary classification was performed using a support vector machine (SVM). TEOAE signal parameters, including signal energy and group delay, were compared between improved (PTA improvement: ≥15 dB) and nonimproved groups using Welch's t-test.
**Results:** Signal energy did not differ ($p = 0.64$) but a significant difference in 1-kHz ($p = 0.045$) group delay was recorded between improved and nonimproved groups. The SVM achieved a cross-validated accuracy of >80% in predicting hearing outcomes.

**Conclusions and Significance:** This study revealed that baseline TEOAE parameters obtained during acute MD episodes, when processed through machine learning technology, may provide information on outer hair cell function to predict hearing recovery.

## Introduction

Ménière's disease (MD) is a dysfunction of the inner ear presenting fluctuating hearing loss (often at low frequencies), tinnitus, vertigo, and the feeling of fullness in the ear. The cause of MD remains unclear, but temporal bone examination of MD-affected human cochlea often reveals enlarged volumes of the endolymphatic space [1]. Thus, endolymphatic hydrops (ELH) has been accepted as a common reason for the overall MD symptoms. For hearing impairment in particular, increase in endolymphatic pressure may cause static mechanical deformation of the organ of Corti (OoC), degeneration of the stria vascularis, damage of hair cells and their stereocilia, and disturbance of basilar membrane motion, and thus affect OoC normal functioning. Documented hearing loss is a criterion necessary for establishing the diagnosis of definite or probable MD. The hearing loss level is also used to stage MD.

Situated within the OoC, outer hair cells (OHCs) are specialized for sensing vibrations and producing reverse electromechanical transduction [2]. Consequently, variation in the electrical potential across the OHC membrane can be converted to an oscillating feedback force to the basilar membrane (BM). If the feedback force has an appropriate synchronization with the wave that propagates along the BM, it can then amplify the wave in a cycle-by-cycle manner [3]. Animal experiments have revealed that OHC-based *cochlear amplifiers* can provide up to a 100–1000x gain at the tone's best place [3], but the gain is compromised when OHCs operate under pathological conditions [4].

*Otoacoustic emissions* (OAEs) are sounds produced inside the cochlea and can be recorded in the ear canal with sensitive microphones [5]. Traditionally, OAEs were categorized as whether or not emission was externally elicited and whether emission was generated as a response through nonlinear distortion or linear reflection [6]. Two main ways to elicit OAEs are in common use: OAEs elicited by two sustained tones are called distortion-product (DP) OAEs, and those elicited by short stimuli are called transient-evoked (TE) OAEs. Both DPOAEs and TEOAEs rely on the normal functioning of OHCs to provide the amplification necessary for them to be detected in the ear canal. Thus, OAEs are clinically useful for testing OHC functions, even though there may be alternative mechanisms to OHC dysfunction that TEOAEs cannot detect.

Theoretically, hearing loss from MD is a reflection of OHC dysfunction. Because of the OoC deformation caused by ELH, OHCs would be driven away from their normal resting positions [7]. This could reduce their ability to provide the feedback force necessary for wave amplification, and hence cause hearing loss with various audiometric profiles.

Fluctuation of hearing impairment is characteristic of MD, and the literature has suggested that hearing loss is a major source of anxiety and depression in MD patients [8,9]. Hearing recovery hence becomes a reason for consultation visits. Currently, there is no way of predicting hearing recovery following acute episodes of MD, although middle- to high-frequency hearing loss is generally considered a poor prognosis sign in MD [10]. It is crucial to understand whether OAE signals can provide useful information to predict MD hearing outcomes for patient-counselling purposes. In this study, we used machine learning technology to analyse OAE signals and established a predictive model for hearing recovery in MD patients.

## Materials and methods

### *Study design*

A total of 30 unilateral MD patients were recruited prospectively upon onset of acute cochleo-vestibular symptoms in a consecutive fashion. All patients met the 1995 AAO-HNS criteria of definite or probable MD. No patients showed retrocochlear signs in auditory brainstem-evoked potential responses. Serial TEOAE and pure-tone audiogram (PTA) data were recorded longitudinally for up to 6 months (visit at index, 1, 2, 3, and 4 weeks; 2, 3, and 6 months). Linear transformations of denoised TEOAE signals between 2.5 and 20 ms were performed to project onto the three most prominent principal directions (principal components PC1, PC2, and PC3). In addition, TEOAE signal parameters, including *total energy* and *group delay* at 1 and 2 kHz, respectively, were compared between *improved* (PTA improvement from index to the last visits: ≥15 dB at any of the following frequencies: 500, 1000, 2000, or 3000 Hz) and *nonimproved* groups using Welch's *t*-test. The Institutional Review Board of Cathay General Hospital approved this study (Approval Number: CGH-P104045). All data analysis steps were carried out in Matlab R2018b with the provided libraries.

### *Data input*

The projection of signals into prominent directions involved a change-of-coordinate process called *principal component analysis* (PCA)[11]. These prominent directions were *orthogonal* in the linear algebraic sense, meaning that they were mutually perpendicular like the x, y, and z axes in Cartesian coordinates.

The principal components were determined by joint consideration of all data, regardless of whether or not they are derived from improved ears, in the following steps. First, denoting the length of TEOAE signals as $L$, a cross-correlation matrix $C$ was computed by defining its element at the $i^{th}$ row and $j^{th}$ column as $C_{ij} = E[(X_i - \mu_i)(X_j - \mu_j)]$, $1 \leq i, j \leq L$, where $E[\cdot]$ stands for the expected value calculated across all ears, $X_i$ and $X_j$ denote the TEOAE signal at time index *i* and *j*, respectively, and $\mu_i$ and $\mu_j$ denote the mean of $X_i$ and $X_j$ across all ears, respectively. Then, by standard matrix diagonalization, an $L \times L$ orthogonal matrix $Q$ could be found such that $CQ = QD$, where $D = \text{diag}(\lambda_1, \lambda_2, \ldots, \lambda_L)$ is a diagonal matrix. Without loss of generality, the order of *eigenvalues* could be chosen such that $\lambda_1 \geq \lambda_2 \geq \cdots \geq \lambda_L \geq 0$. Finally, the column vectors of $Q$ would be the principal components, in decreasing order of prominence and all orthogonal to one another.

Once these orthogonal directions were determined, any TEOAE signal could be mapped from its waveform to a point in a 3-dimensional (3D) space by projecting onto the first three column vectors of $Q$. The result is denoted as ***x*** = (PC1, PC2, PC3) hereafter. In this study, PCA enabled data visualization and ensured that subsequent machine learning procedures do not overfit the data to maximize prognosis accuracy.

### *Prognosis by machine learning*

The support vector machine (SVM)[12] was adopted and *trained* to predict prognosis. The training process consisted of two key steps: first, the machine required *supervision* in which the data needed to be labelled as improved or nonimproved. Second, potentially crucial features were defined based on domain knowledge. In this study, PC1, PC2, PC3; signal energy; and group delay at low frequencies (1.0 and 2.0 kHz) were selected as crucial features. Afterwards, the SVM automatically computed a ($n$-1) dimensional hyperplane or hypersurface that optimally separated the data in an $n$-dimensional space, $n$ being the number of features. Figure 1 illustrates the separation of the improved and nonimproved ears in the 3D space of (PC1, PC2, PC3).

### *Clinical relevance of selected features*

*Group delay* (GD) refers to the time each frequency component appears in a signal. Whereas PC1, PC2, and PC3 captured the hidden features in the OAE signals that may not have been of clinical relevance directly, signal energy and the TEOAE GD could reflect cochlear function to an extent. Signal energy was defined as the sum of the square of the signal over a period (2.5–20 ms) and it captured signal strength. The physical meaning of GD is illustrated in Figure 2. In the literature of OAEs, GD has been calculated through Fourier transform by estimating the first derivative of the unwrapped phase spectrum [13].

### *Evaluation of SVM performance*

To evaluate the usefulness of the SVM in providing accurate prognosis, *cross-validation* was applied. Data were first separated into K piles (K = 5 for example), and each pile of data would take turns to be the *validation set* while other piles were used as the *training set*. The SVM learnt the separation hyperplane from the training set, and its performance was evaluated using the validation set. Finally, the accuracy was averaged across the K piles of data, and the process was referred to as *K-fold validation*.

### *Details regarding signal acquisition*

TEOAE signals were recorded using the ER-10C (Etymotic Research Inc., Elkgrove

Village, IL, USA) probe connected to a 24-bit UltraLite-mk3 Hybrid external sound card (MOTU, Cambridge, MA, USA). The transient click was generated by a computer and transmitted through the soundcard to the ER-10C earphone. The sampling rate was set to 44.1 kHz. The peak sound pressure level was approximately 74 dB SPL (relative to 20 µPa). The click rate was maintained at 10 per second. The TEOAE was recorded at Cathay General Hospital, Taipei, and the noise floor was approximately 23–27 dB SPL. For each recording session, 3000 clicks were played and responses with artefacts (such as the swallowing noise) were rejected. Afterwards, noise level was reduced by taking the samplewise median to estimate the true signal. Both the MD-affected ear and the opposite ear were measured. A typical example of an estimated TEOAE signal and the noise standard deviation is presented in Figure 3a. This Fourier-transform magnitude of this signal is also shown in Figure 3b.

# Results

### *Statistical analysis of parameters*

All participants had no vestibular symptom at the end of follow-up. Table 1 lists the results of analysing the difference between hearing improved (N=14) and non-improved (N=16) ears on the basis of Welch's *t*-test with unequal variances. The energy parameter did not reveal a significant difference between the two groups, even though the mean among improved ears, $5.57 \times 10^{-9}$ Pa$^2$, was indeed higher than that of nonimproved ears, $4.65 \times 10^{-9}$ Pa$^2$. We discovered that the distribution GD at 1 kHz was significantly different between improved and nonimproved ears. Examples of TEOAE signals with long and short GD, respectively, are shown in Fig. 4; the one with longer GD was obtained from the improved group, and the one with shorter GD was obtained from the nonimproved group.

### *SVM performance in prognosis prediction*

Table 2 evaluates SVM performance by 5-fold cross validation. In this study, the sigmoid kernel function was selected for configuring the SVM. Because SVM performance depended on the penalty parameter *C* and the sigmoid kernel parameter *γ*, Table 2 only reports the results of using the parameter values (*C, γ*) in the range of $2^{-6}$ to $2^6$ that produced the highest prognosis accuracy. These optimal (*C, γ*) values are also listed in Table 2.

    By using the PCs or GD at 1 or 2 kHz, the SVM could achieve >80% prognosis accuracy based on TEOAE data that were recorded during patients' first visits to the clinic. By contrast, the prognosis accuracy provided by the energy parameter alone was approximately 71%.
    To summarize, in the present study, we recruited MD patients at the onset of acute sensorineural hearing loss (ASNHL). We used paired PTA and TEOAE to follow hearing recovery longitudinally. By applying machine learning techniques, we demonstrated that, first, several useful parameters emerged through PCA (PC1, PC2, and PC3) and additional relevant parameters (such as GD at 1 and 2 kHz) could also be

included based on domain knowledge. Second, the optimal way of combining parameters could be computed through the SVM to maximize prognosis accuracy and prevent data overfitting.

## Discussion

ASNHL is a critical ear condition that requires immediate medical attention. ASNHL may result from MD, sudden deafness, perilymph fistula, or rare autoimmune diseases. Its *acute* and *fluctuating* nature is characteristic of MD, and hearing outcome of MD following acute attacks should be the focus of otology consultations. However, currently there is no way of predicting hearing recovery for acute MD episodes, causing much frustration and anxiety to patients and physicians [8,9].

The OHC is the critical function unit of the cochlea, and its output reflects the effective hearing threshold. Damage to OHCs may cause temporary or permanent functional deficits and exhibit sensory hearing impairment clinically. OAE is a unique and commonly used tool for detecting OHC functions. No other clinical test batteries can reflect the function of other individual cochlear organs.

OAEs have been studied extensively for diagnosis or hearing monitoring purposes. For instance, the TEOAE was still detectable in MD-affected ears [15]. The pattern of DPOAE suppression with low-frequency tones (<100 Hz) is different between MD and control ears with non-Ménière's ASHNL [16]. Recently, Drexl et al. [7] compared DPOAE in MD ears, nonaffected ears in MD patients, and ears in people with normal hearing. In addition to the commonly used cubic distortion component of $2f_1-f_2$, they also measured quadratic distortion. Cubic distortion was significantly reduced in MD ears; however, quadratic distortion remained relatively unaffected.

Several studies have attempted to use OAEs for idiopathic ASNHL prognosis [17,18,19]. In particular, Shupak et al. [19] demonstrated that both the TEOAE and DPOAE, when measured at most affected frequencies, could predict idiopathic ASNHL outcome with >70% and >80% sensitivity, respectively, at 100% specificity.

Although several studies have attempted this, few have used OAEs for MD hearing prognosis. Since hearing loss in MD originates from OHC malfunction, OAE should carry certain hidden signals that pertain to chances of hearing recovery. The present study demonstrated a method of identifying such hidden signals in the TEOAE through data-driven computational approaches.

The current finding on GD may have indicated that the improved ears had more sharply tuned frequency responses inside the cochlea. The GD of OAEs from the hearing organ across many species has been found to be positively correlated to cochlear tuning sharpness [14], which would be compromised when the OHC is damaged. Interestingly, in the present study, we also analysed GD at higher frequencies (e.g., 4 kHz); however, no significant difference was revealed between improved and nonimproved ears.

Our findings that the TEOAE GD at 1 and 2 kHz became shorter in nonimproved ears corresponded with the results of a previous report by Takeda et al. [17]. In their work, a 'slow component' (i.e., a component that occurred late) of the TEOAE at approximately 1.0 kHz was more likely to be absent in MD ears with ≥30 dB HL than in those with <30 dB HL. Mathematically, missing a late component would move the centroid of the signal toward an earlier time and shorten GD; thus, Takeda et al.'s findings corresponded with the results of this research.

The DPOAE could be equally useful in providing information for MD prognosis. Indeed, the DPOAE level is well known to be negatively correlated to the degree of hearing loss; however, the TEOAE allows observation of the entire waveform and analysis of its change over time. Recently, techniques for deciphering DPOAE dynamics over time have also been developed [20], so its application for predicting hearing recovery may be worthy of further investigation.

## Conclusion

This study revealed that baseline TEOAE features (GD extracted at 1 and 2 kHz and component values PC1, PC2, and PC3) during an acute MD episode, when processed with machine learning technology, could provide information regarding OHC function for predicting hearing recovery.

## Acknowledgement

Authors thank Mr. Daniel Wu's help to organize the literatures.

## Funding

**Table 1:** Results of parameter distribution.

| Parameter | $t$-value | $p$-value | Mean±SD: improved | Mean±SD: Non-improved |
|---|---|---|---|---|
| PC1 | 1.56 | 0.136 | -- | -- |
| PC2 | −1.10 | 0.280 | -- | -- |
| PC3 | 2.04 | 0.057 | -- | -- |
| Energy | 0.48 | 0.636 | $(5.57 \pm 5.40) \times 10^{-9}$ | $(4.65 \pm 4.73) \times 10^{-9}$ |
| GD at 1.0 kHz | 2.11 | **0.045** | 4.29±1.18 | 3.43±0.95 |
| GD at 2.0 kHz | 2.04 | 0.053 | 3.75±1.03 | 3.05±0.72 |

In the second column, a positive $t$-value ($t > 0$) means that the parameter in improved ears has a larger sample mean than the non-improved ears. The corresponding $p$-values are listed in the third column, where the bold font marks those parameters with $p < 0.05$, a common criterion for statistical significance. The mean and standard deviation (SD) for energy and GD at both frequencies are also shown for the improved and non-improved groups, separately. The unit of energy is Pa$^2$, and the unit for GD is ms.

**Table 2:** Prognosis accuracy by different features.

| Parameters | Accuracy (%) | C | γ |
|---|---|---|---|
| (PC1, PC2, PC3) | 82.7 | 11.3 | 2.46 |
| Energy | 70.8 | 64 | 19.7 |
| GD at 1 kHz | 84.1 | 64 | 64 |
| GD at 2 kHz | 80.8 | 64 | 19.7 |

Results were obtained by 5-fold validation using the best combination of SVM penalty parameter *C* and sigmoid kernel parameter *γ* shown in the last two columns.

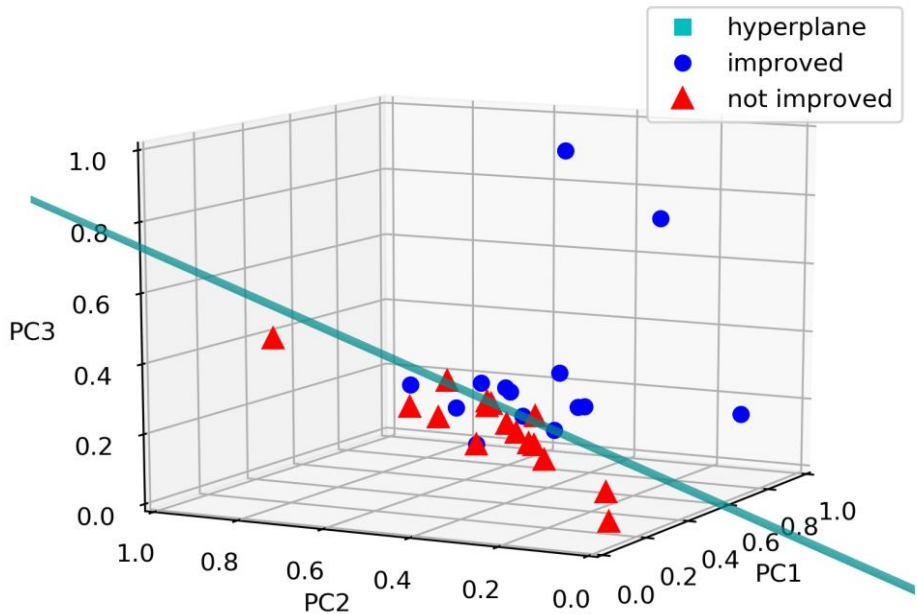

**Figure 1.** Illustration of support vector machine (SVM). The data points here are derived from actual TEOAE waveforms by projection onto PC1, PC2, and PC3. The thick straight line depicts the plane that optimally separates the two groups. If the data between two groups were completely separable, all the dots (improved) would have been above the plane and the triangles (not improved) would have been below; the accuracy of prognosis would have been 100%.

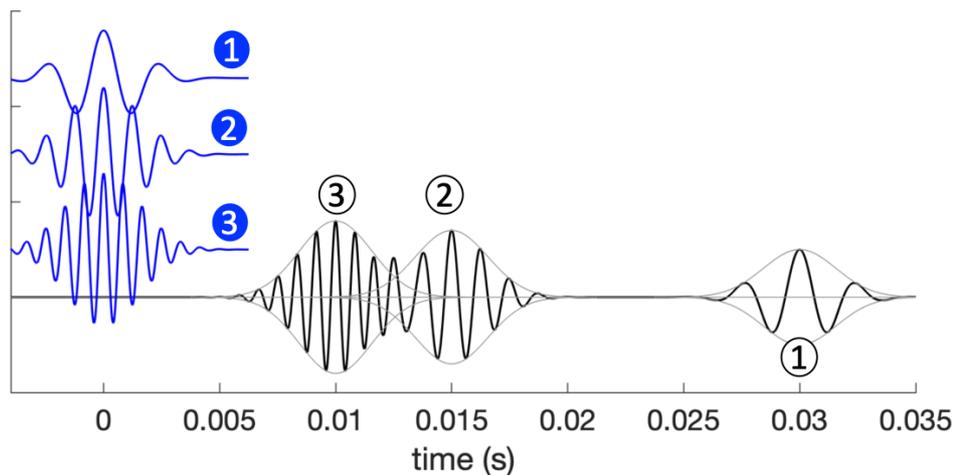

**Figure 2.** (color online) *Group delay* (GD) describes the latency of waves traveling with dispersion; that is, different frequencies may travel at different speeds. In this hypothetical example, three wave packets enter the cochlea simultaneously at $t = 0$ but come out with GD = 0.030, 0.015, and 0.010 s, respectively (black). The thin line marks the signal envelope for bettering viewing.

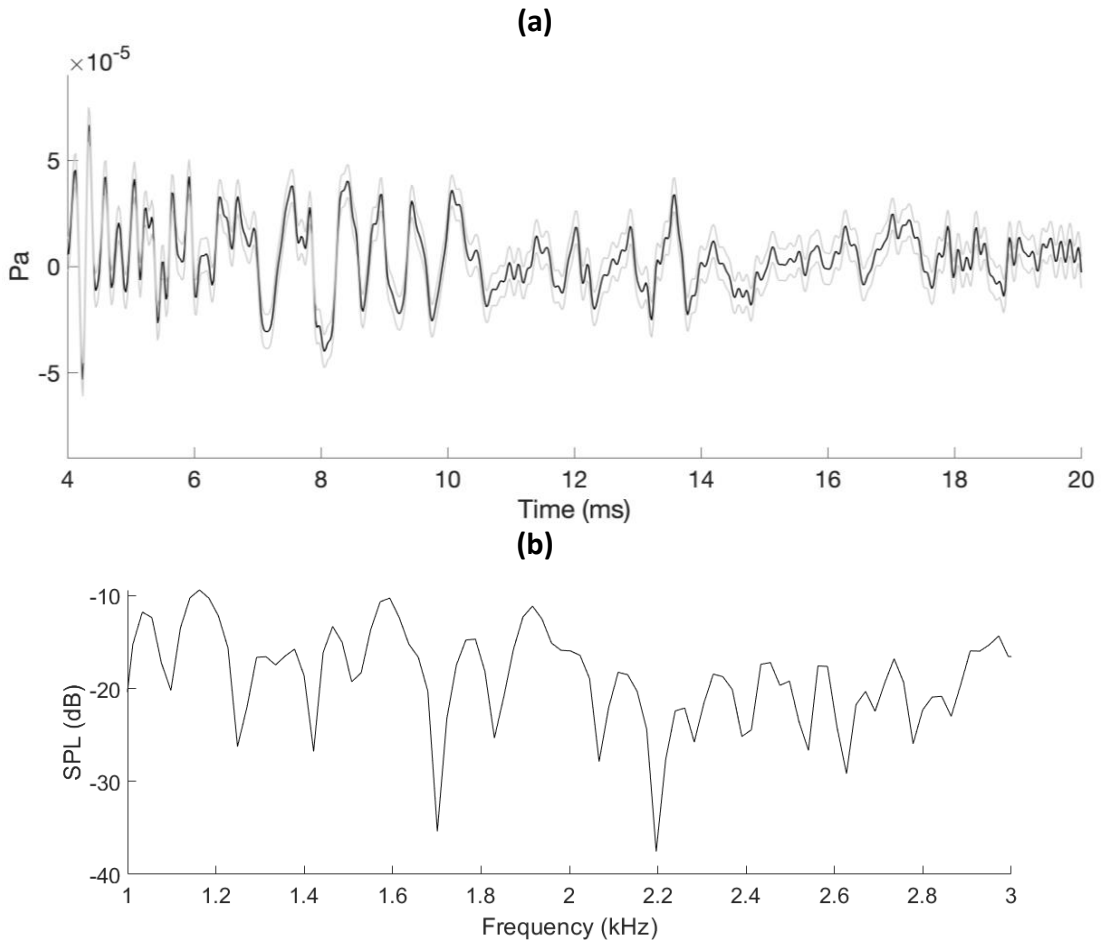

**Figure 3.** A typical TEOAE signal and the corresponding spectrum acquired from a non-MD ear. 3000 clicks were played and data were subject to artefact rejection and then averaged to improve the signal to noise ratio. **(a)** The dark line shows the average and the gray lines above and below show the signal $\pm 1$ standard deviation of the noise, which was estimated after signal averaging. **(b)** The magnitude spectrum of this signal in dB scale, plotted from 1 to 3 kHz.

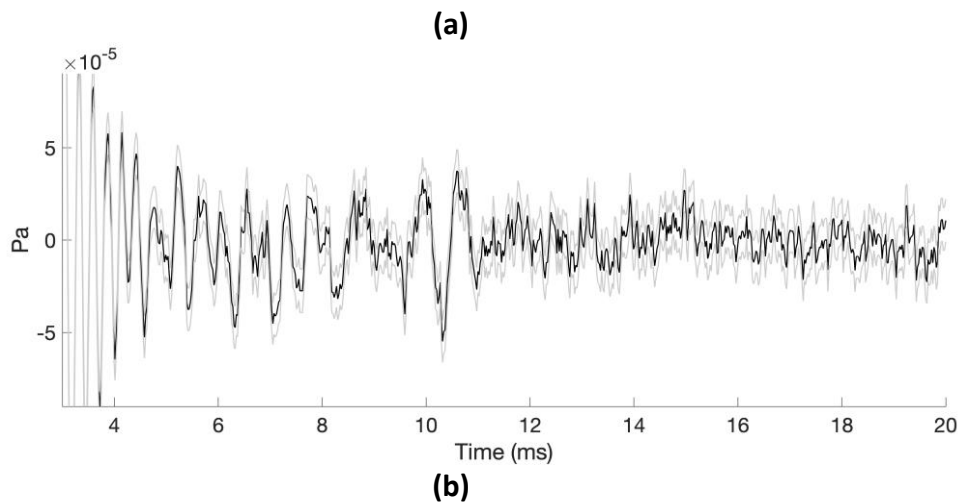

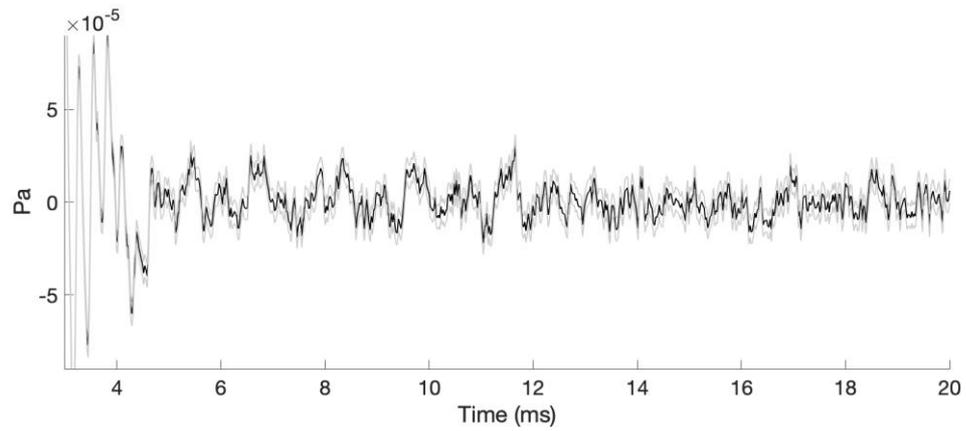

**Figure 4.** Examples of TEOAE traces. (a) This example was obtained from an improved ear and it had relatively long GDs (6.10 and 5.09 ms at 1 kHz and 2 kHz, respectively). (b) This example was obtained from a nonimproved ear and the GDs were relatively short (3.46 and 3.19 ms at 1 kHz and 2 kHz, respectively**)**.